# Thermoelectric properties of graphyne from first-principles calculations

P. H. Jiang, H. J. Liu[*], L. Cheng, D. D. Fan, J. Zhang, J. Wei, J. H. Liang, J. Shi

*Key Laboratory of Artificial Micro- and Nano-Structures of Ministry of Education and School of Physics and Technology, Wuhan University, Wuhan 430072, China*

The two-dimensional graphene-like carbon allotrope, graphyne, has been recently fabricated and exhibits many interesting electronic properties. In this work, we investigate the thermoelectric properties of $\gamma$-graphyne by performing first-principles calculations combined with Boltzmann transport theory for both electron and phonon. The carrier relaxation time is accurately evaluated from the ultra-dense electron-phonon coupling matrix elements calculated by adopting the density functional perturbation theory and Wannier interpolation, rather than the generally used deformation potential theory which only considers the electron-acoustic phonon scattering. It is found that the thermoelectric performance of $\gamma$-graphyne exhibits a strong dependence on the temperature and carrier type. At an intermediate temperature of 600 K, a maximum *ZT* value of 1.5 and 1.0 can be achieved for the *p*- and *n*-type systems, respectively.

## 1. Introduction

The steady increase in the world's population and its demands for fuel and products cause energy crisis in the past 50 years. Moreover, many industrial and commercial energy utilizations result in excessive rates of waste heat rejection. The thermoelectric technology is believed to be one of the effective methods for energy harvesting since it provides a promising route to convert waste heat into electricity. The efficiency of a thermoelectric material is determined by the dimensionless figure-of-merit $ZT = S^2 \sigma T / (\kappa_e + \kappa_{ph})$, where $S$, $\sigma$, $T$, $\kappa_e$ and $\kappa_{ph}$ are the Seebeck coefficient, the electrical conductivity, the absolute temperature, the electronic and phonon thermal conductivity, respectively. Good thermoelectric material has larger *ZT* value and one therefore must try to maximize the power factor ($S^2\sigma$) and/or minimize the thermal conductivity ($\kappa_e + \kappa_{ph}$). However, it is extremely difficult to do so since these transport coefficients are usually coupled with each other in conventional thermoelectric materials [1]. In recent years, the successful fabrication of low-dimensional thermoelectric materials has simulated a lot of research interest [2, 3, 4, 5] because the *ZT* value can be enhanced remarkably due to the quantum confinement effect [6, 7]. On the other hand, it is highly desired that better thermoelectric performance could be realized in the earth-abundant and environment-friendly systems, e.g., carbon materials. In this respect, the two-dimensional graphene seems to be a possible choice since its first fabrication in 2004 [8]. The existence of Dirac-cone band structure makes graphene exhibit numerous novel electronic properties [9]. However, the absence of band gap leads to very smaller Seebeck coefficient of graphene. Together with extraordinarily high thermal conductivity, the thermoelectric performance of graphene is indeed extremely poor [10].

---

[*] Corresponding author. E-mail: phlhj@whu.edu.cn (H. J. Liu)

Another two-dimensional candidate in the carbon family is graphyne, which was first proposed theoretically by Baughman *et al.* in 1987 [11]. It can be viewed as modified graphene by inserting the carbon-carbon triple bonds (*sp* hybridization) into the $sp^2$ hybridized graphene. A series of atomic structures, e.g., the *α*-, *β*-, *γ*-, 6, 6, 12-graphyne and graphdiyne, can be obtained by varying the number and position of the triple bonds [11, 12]. Recently, the successful fabrication of large area graphyne films [13] has inspired extensive studies exploring its mechanical, thermal, electronic and optical properties [14, 15, 16, 17]. Compared with graphene, graphyne exhibits more amazing electronic properties since the Dirac cones with different symmetries are presented [15]. Moreover, the band gap is opened up in the *γ*-graphyne and graphdiyne [18, 19]. Such novel characteristics extend the application prospects of the two-dimensional carbon allotropes. The presence of a band gap can drastically increase the Seebeck coefficient [20], and the inserted triple bonds can reduce the thermal conductivity significantly [19, 21, 22]. All these observations suggest that the graphyne systems with finite band gap could exhibit very favorable thermoelectric performance which deserves a complete understanding.

In this work, the thermoelectric properties of the semiconducting γ-graphyne is systematically investigated by using first-principles calculations and Boltzmann transport theory, where the carrier relaxation time is accurately evaluated from the ultra-dense electron-phonon coupling matrix elements. We demonstrate that the thermoelectric performance of γ-graphyne exhibits a marked dependence on the temperature and carrier type. At an intermediate temperature of 600 K, a maximum *ZT* value of 1.5 and 1.0 can be respectively achieved for the *p*- and *n*-type systems, which suggests that good thermoelectric performance can be also achieved in previously unexpected carbon systems.

## 2. Computational methods

Our first-principles total energy calculations are performed within the framework of density functional theory (DFT), as implemented in the QUANTUM ESPRESSO package [23]. We use the norm-conserving pseudopotential and the exchange-correlation functional is in the form of Perdew-Burke-Ernzerhof [24]. The system is modeled by adopting a hexagonal supercell geometry where the vacuum distance is set to 14 Å to eliminate the interactions between the graphyne layer and its periodic images. The kinetic energy cutoff is 80 Ry for the wavefunction and 800 Ry for the charge density. For the phonon dispersion relations and the electron-phonon coupling matrix elements, we apply the density functional perturbation theory (DFPT) [25] and Wannier interpolation technique [26]. The calculations are initially done by using a coarse 3×3×1 **q** and **k** mesh, and then interpolate to a dense mesh of 120×120×1 via the maximally localized Wannier functions as implemented in the electron-phonon Wannier (EPW) package [27]. After obtaining the electron self-energy $\Sigma_{n\mathbf{k}}$ for band *n* and state **k** from the interpolated ultra-dense electron-phonon coupling matrix elements, the relaxation time can be readily determined by $\left(\tau_{n\mathbf{k}}\right)^{-1} = 2\left[\text{Im}\left(\Sigma_{n\mathbf{k}}\right)\right]/\hbar$ [28], where $\hbar$ is the reduced Plank constant.

Based on the energy band structure and carrier relaxation time, the electronic transport coefficients can be calculated by using the following formulas as derived from Boltzmann theory [29]:

$$S = -\frac{1}{eT} \frac{\sum_{n,\mathbf{k}} \left(E_{n\mathbf{k}} - E_f\right) v_{n\mathbf{k}}^2 \tau_{n\mathbf{k}} \frac{\partial f_{n\mathbf{k}}}{\partial E_{n\mathbf{k}}}}{\sum_{n,\mathbf{k}} v_{n\mathbf{k}}^2 \tau_{n\mathbf{k}} \frac{\partial f_{n\mathbf{k}}}{\partial E_{n\mathbf{k}}}}, \quad (1)$$

$$\sigma = \frac{1}{NV} \sum_{n,\mathbf{k}} -e^2 v_{n\mathbf{k}}^2 \tau_{n\mathbf{k}} \frac{\partial f_{n\mathbf{k}}}{\partial E_{n\mathbf{k}}}, \quad (2)$$

Here $E_{n\mathbf{k}}$ is the energy eigenvalue, $E_f$ is the fermi energy, $v_{n\mathbf{k}}$ is the group velocity, $\tau_{n\mathbf{k}}$ is the relaxation time, $f_{n\mathbf{k}}$ is the Fermi occupation, $N$ is the total number of **k** points, and $V$ is the volume of the primitive cell (with respect to a vacuum distance of 3.35 Å). The electronic thermal conductivity $\kappa_e$ is derived from the electrical conductivity $\sigma$ according to the Wiedemann-Franz Law $\kappa_e = L\sigma T$ [30], where the Lorenz number $L$ for the two-dimensional system is expressed as [6]:

$$L = \frac{\kappa_e}{\sigma T} = \left(\frac{k_B}{e}\right)^2 \left[\frac{3F_2}{F_0} - \left(\frac{2F_1}{F_0}\right)^2\right]. \quad (3)$$

with the Fermi integral $F_i = F_i(\eta) = \int_0^\infty \frac{x^i dx}{e^{(x-\eta)}+1}$ ($\eta$ is the reduced Fermi energy).

The phonon thermal conductivity $\kappa_{ph}$ can be obtained by solving the phonon Boltzmann transport equation as implemented in the so-called ShengBTE code [31], where the second-order and third-order interatomic force constants are calculated with a 4×4×1 supercell. The interactions up to the fourth nearest neighbors are included for the anharmonic one, and a **q** point grid of 34×34×1 is chosen to ensure the convergence of $\kappa_{ph}$.

## 3. Results and discussion

The crystal structure of $\gamma$-graphyne is displayed in Figure 1, with 12 carbon atoms included in the primitive cell. Such a graphene-like structure can be viewed as carbon hexagons connected by the carbon-carbon triple bonds, so that the same symmetry (*P*6/*mmm*) of graphene is maintained. The optimized lattice constants are $a = b = 6.890$ Å, and three different bond lengths exist due to the mixed hybridization of carbon atoms with $sp^2$–$sp^2$ (1.426 Å), $sp^2$–$sp$ (1.408 Å), and $sp$–$sp$ (1.223 Å). These structure parameters are in good agreement with previously calculated using projector-augmented-wave (PAW) approach [12, 19]. To check the stability of the structure, we have calculated the phonon dispersion relations of $\gamma$-graphyne and no imaginary frequency is found, as shown in Figure 2(a). In Fig. 2(b), we plot the band structure of $\gamma$-graphyne. Unlike graphene with a Dirac cone, we see a direct band gap of 0.46 eV opened up due to the presence of $sp$ hybridization. Both the valence band maximum (VBM) and the conduction band minimum (CBM) are located at the M point of the Brillouin zone, instead of the K point for graphene. Our calculated band structure is consistent with previous theoretical studies using PAW [19] and pseudopotential methods [20].

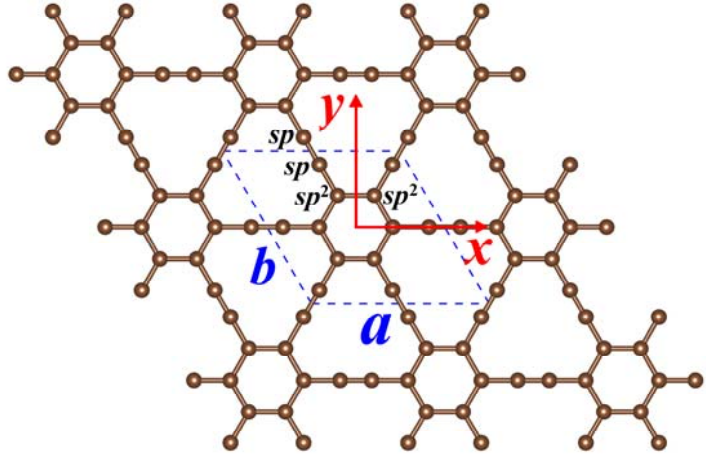

**Figure 1** The atomic structure of $\gamma$-graphyne. The blue dashed lines indicate the primitive cell with basis vectors *a* and *b*. The conventional *x*- and *y*-axis are also indicated.

When dealing with the electronic transport properties, the carrier relaxation time should be carefully treated. Earlier attempts in addressing this fundamentally important issue either adopt the ballistic transport model without consideration of the scattering [20, 21], or apply deformation potential (DP) theory which only considers the electron-acoustic phonon scattering [19, 32]. In the present work, the carrier relaxation time is accurately evaluated by considering the complete electron-phonon coupling within the EPW framework. It should be emphasized that a very dense **k** and **q** mesh should be used to obtain the electron-phonon coupling matrix elements, which can be done by adopting the Wannier interpolation technique based on the DFT and DFPT calculations. In Fig. 2(c) and 2(d), we respectively plot the interpolated phonon dispersion relations and the electronic band structure, which agree well with those obtained directly from DFPT and DFT approaches and indicate the accuracy and convergence of the Wannier interpolation. Figure 3(a) shows the calculated electron-phonon scattering rate (reciprocal of the relaxation time $\tau_{n\mathbf{k}}$) and the electronic density of states (DOS) as a function of energy. We see that the scattering rate is proportional to the DOS, which is expected since the DOS reflects the phase space available for carrier scattering [33, 34]. In addition, the scattering rate increases with increasing temperature, indicating that the relaxation time is smaller at higher temperature. In most of previous works, the DP theory is generally used to predict the relaxation time [19, 35, 36, 37] for the VBM and CBM states. For example, the room temperature relaxation times of *γ*-graphyne given by DP theory are $4.9 \times 10^{-13}$ s and $14.1 \times 10^{-13}$ s for the *p*-type (VBM) and *n*-type (CBM) systems, respectively [19]. In contrast, our relaxation time obtained from the EPW method is energy dependent ($\tau_{n\mathbf{k}}$). For comparison, we depict in Fig. 3(b) the temperature dependent relaxation times for the VBM and CBM states. It is clear to find that our results are obviously lower than those predicted from the simple DP theory. The reason is that only the acoustic phonon scattering is considered in the DP theory [38] while the scattering from all the phonon modes is included in our EPW approach. Such difference also suggests that the optical phonon scattering could play an important role in determining the relaxation time and cannot be neglected in the γ-graphyne system. The overestimated relaxation time in previous work may lead to an unexpected high thermoelectric performance [19], as will be discussed later.

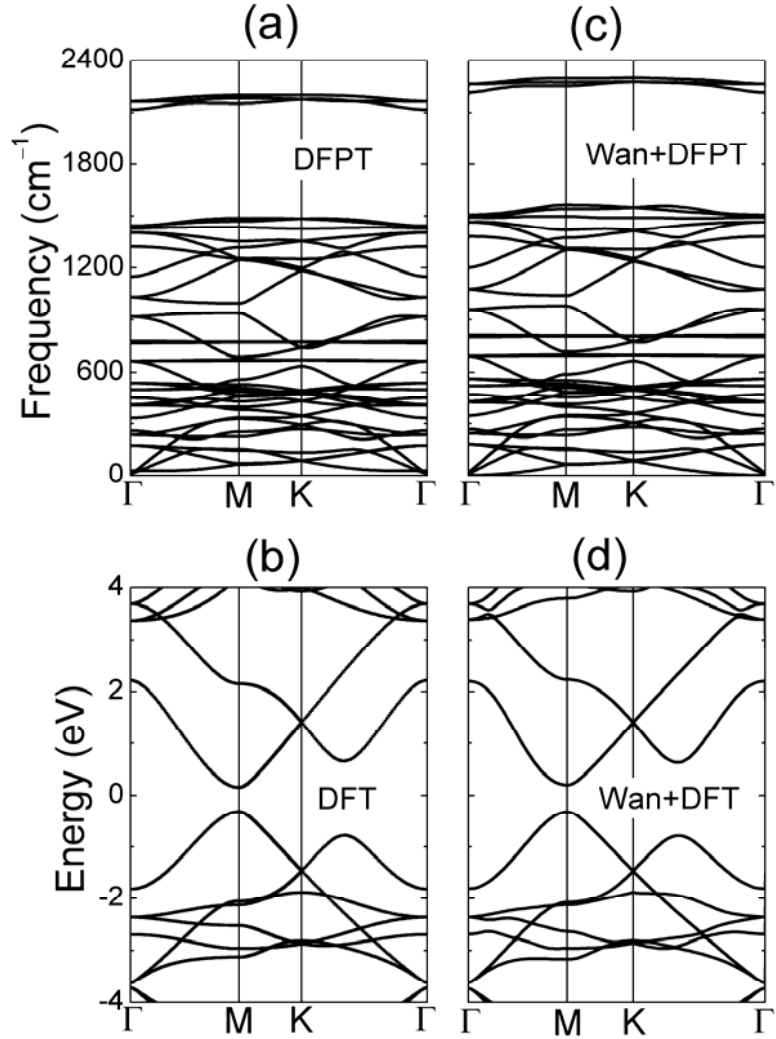

**Figure 2** The phonon dispersion relations and electronic band structures of γ-graphyne calculated by using (a) DFPT, (b) DFT, (c) Wannier interpolations based on DFPT (Wan+DFPT), and (d) Wannier interpolations based on DFT (Wan+DFT).

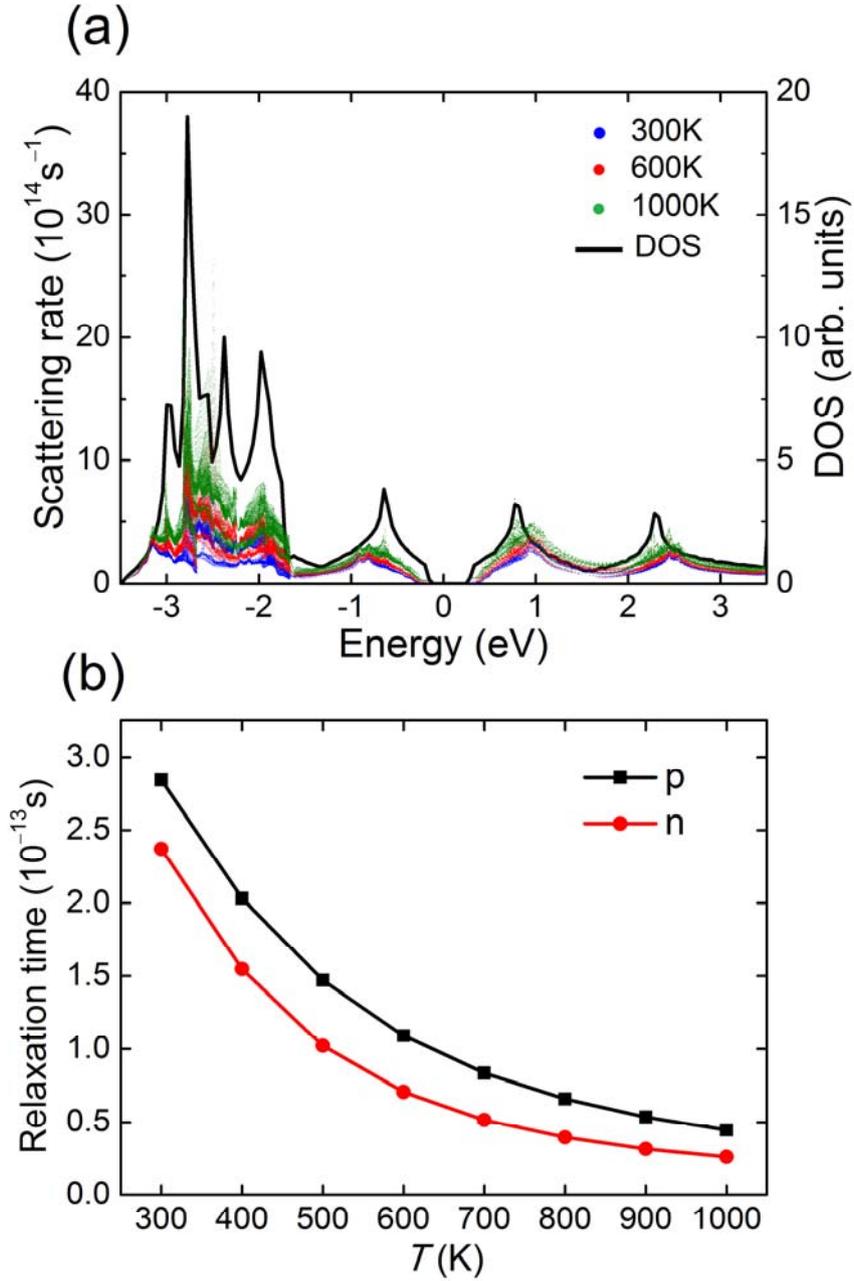

**Figure 3** (a) The electron-phonon scattering rate of $\gamma$-graphyne at 300 K, 600 K and 1000 K (left), and the corresponding density of states (right). (b) The temperature dependent relaxation time of the energy states at VBM and CBM.

The electronic transport coefficients of $\gamma$-graphyne can be now evaluated from Eq. (1) and (2) by inserting the energy dependent relaxation time. The room temperature Seebeck coefficient $S$, the electrical conductivity $\sigma$, and the power factor $S^2\sigma$ are plotted in Figure 4 as a function of carrier concentration. From the inset of Fig. 4(a), we can see that the Seebeck coefficient exhibits peak values at very low carrier concentration, and the absolute values can be as high as 670 μV/K and 620 μV/K for the *p*-type and *n*-type systems, respectively. These values are much larger than those of the conventional thermoelectric materials such as $Bi_2Te_3$ [39], which is very beneficial for its thermoelectric performance. With increasing carrier concentration, however, the Seebeck coefficient decreases obviously and becomes vanished when the concentration is larger than $10^{13}$ $cm^{-2}$. On the contrary, the electrical conductivity (Fig. 4(b)) increases sharply when the carrier concentration is larger than $10^{12}$ $cm^{-2}$ but maintains a rather small value at low carrier concentration range where the Seebeck coefficient is large enough. Such an opposite behavior calls for a compromise between the Seebeck coefficient and the electrical conductivity, so that the maximum power factor can be achieved (Fig. 4(c)). At moderate carrier concentrations of $2.39 \times 10^{12}$ $cm^{-2}$ and $1.54 \times 10^{12}$ $cm^{-2}$, the optimized power factors are 0.37 W/mK$^2$ and 0.24 W/mK$^2$ for the *p*-type and *n*-type systems, respectively. In addition, the electronic thermal conductivity $\kappa_e$ is derived from the Wiedemann-Franz Law, where the Lorentz number is $1.2$–$1.4 \times 10^{-8}$ WΩ/K$^2$ calculated from Eq. (3). The electronic thermal conductivity shows similar behavior with the electrical conductivity and is thus not shown here. It should be noted that the transport coefficients, particularly the electrical conductivity, exhibit an obvious dependence on the direction, which can be explained by the anisotropic group velocity of *γ*-graphyne. In Figure 5(a), we plot the group velocity for the highest valence band, which is quite different at different direction. For example, the group velocity along $k_x$ and $k_y$ directions are calculated to be $3.02 \times 10^5$ m/s and $1.31 \times 10^5$ m/s at the M point, respectively. Similar behavior can be found for the lowest conduction band, as indicated in Fig. 5(b). Such anisotropic electronic transport would lead to the direction dependence of the *ZT* value, which will be discussed later.

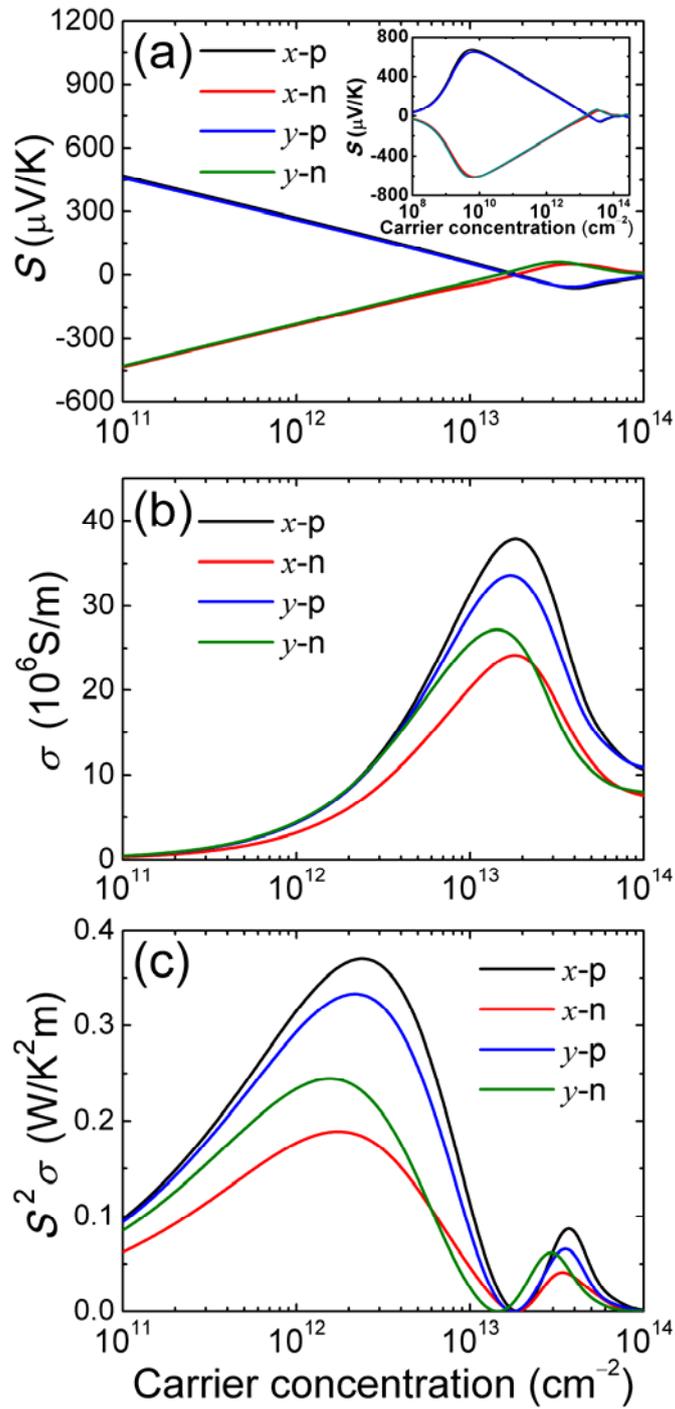

**Figure 4** The room temperature (a) Seebeck coefficient $S$, (b) electrical conductivity $\sigma$, and (c) power factor $S^2\sigma$ of $\gamma$-graphyne as a function of carrier concentration along the $x$- and $y$-directions for both $p$-type and $n$-type systems. The inset of (a) plots the Seebeck coefficient in a large range of carrier concentration to display the peak value.

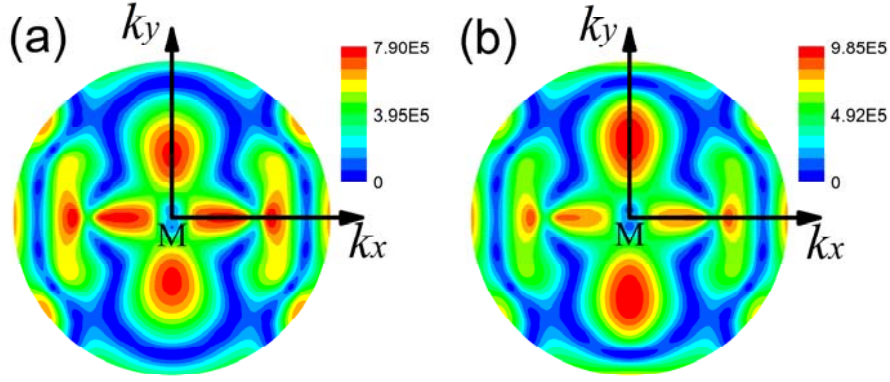

**Figure 5** The group velocity of *γ*-graphyne for (a) the highest valence band, and (b) the lowest conduction band.

We now focus on the phonon transport properties of *γ*-graphyne. As mentioned above, the phonon thermal conductivity $\kappa_{ph}$ can be obtained by solving the phonon Boltzmann transport equation. Figure 6 shows the calculated $\kappa_{ph}$ as a function of temperature from 300 K to 1000 K (left side). We see that the $\kappa_{ph}$ along the *x*- and *y*-directions coincides with each other, and the room temperature value is 76.4 W/mK (calculated with respect to a vacuum distance of 3.35 Å), which is almost two orders of magnitude lower than that of graphene (3080~5150 W/mK in Ref. 40). The significant reduction of $\kappa_{ph}$ mainly results from the existence of the *sp* hybridization of carbon atoms. It was demonstrated that the *sp* bonds in graphyne are weaker than the $sp^2$ bonds in graphene, thus an inefficient heat transfer by lattice vibrations is introduced [22]. On the other hand, the lower atomic density of graphyne is also believed to be an important factor to reduce the phonon induced thermal conductivity [21]. We further find that the $\kappa_{ph}$ is almost inversely proportional to the temperature, revealing that the Umklapp process predominate the phonon scattering in the temperature range considered [41].

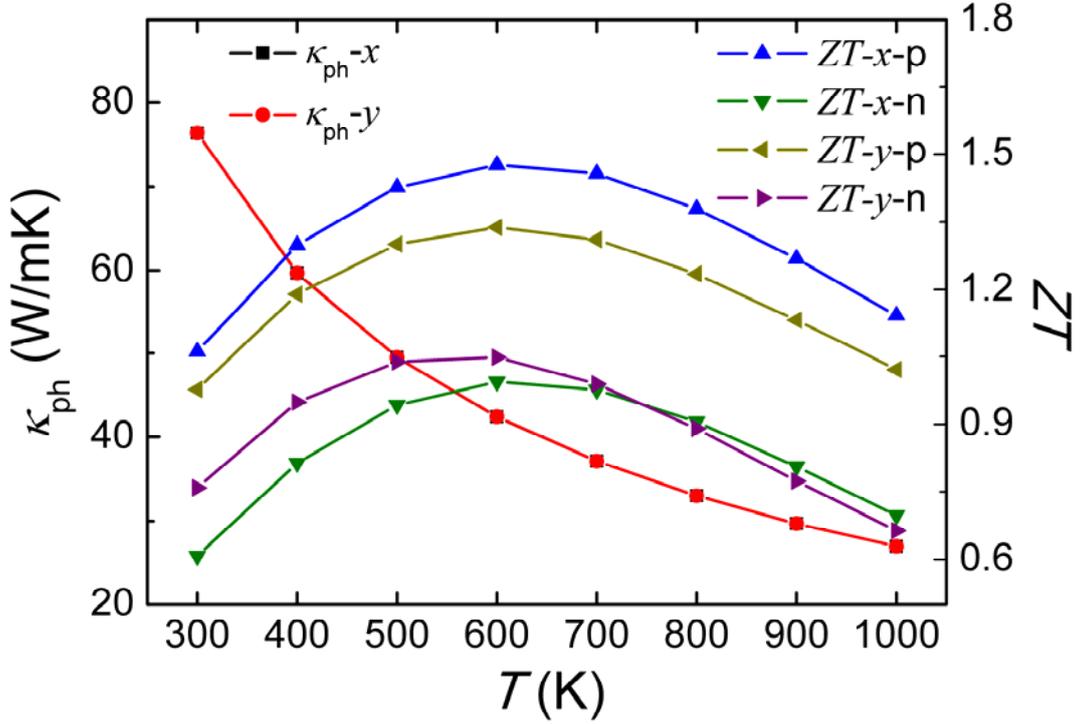

**Figure 6** The temperature dependence of phonon thermal conductivity $\kappa_{ph}$ of $\gamma$-graphyne along the $x$- and $y$-directions (left), and the corresponding $ZT$ values (right).

With all the transport coefficients obtained, we can now evaluate the thermoelectric performance of $\gamma$-graphyne. The right side of Fig. 6 gives the $ZT$ values of $p$-type and $n$-type systems as a function of temperature, where the results for the $x$- and $y$-directions are both shown. In the whole temperature range considered, we see that the $ZT$ values of $p$-type graphyne is much larger than those of the $n$-type system, and this is the case for both the $x$- and $y$-directions. On the other hand, we see that the $ZT$ values exhibit obvious direction dependence, especially for the $p$-type graphyne, which can be attributed to the anisotropic group velocity discussed above. It is interesting to note that regardless of the directions and carrier types, the maximum $ZT$ value always appears at 600 K, which is 1.5 for the $p$-type system along the $x$-direction and 1.0 for the $n$-type system along the $y$-direction. Moreover, we see that in a broad temperature range from 300 K to 1000 K, the $p$-type $ZT$ values are always higher than 1.0 along both directions, which is very desirable in the application of thermoelectric materials. It should be mentioned that the maximum $ZT$ value of 1.5 is much lower than previously predicted result of 2.9 at 760 K [19], which is caused by the overestimation of the relaxation time in that work using the simple DP theory that do not consider the contribution from the optical phonon scattering. On the other hand, our calculated $ZT$ values are significantly larger than that of graphene ($ZT < 0.01$ in Ref. 42), which originates from the increased Seebeck coefficient and reduced thermal conductivity as discussed above. Table I summarizes the optimized $ZT$ values and the corresponding transport coefficients at different temperature. We see that both the power factor and the phonon thermal conductivity decrease with increasing temperature, which could lead to the maximum $ZT$ values at intermediate temperature of 600 K. If we compare the contribution of the thermal conductivity from the electronic and phonon parts, we find that the thermal transport is dominated by the phonon at low temperature. When the temperature becomes higher, the two parts are comparable to each other. All these findings provide useful means to effectively modulate the transport coefficients so that the thermoelectric performance of $\gamma$-graphyne can be further enhanced.

**Table I** Optimized *ZT* values of *p*-type and *n*-type *γ*-graphyne along the *x*- and *y*-directions at different temperature. The corresponding carrier concentration, the transport coefficients, and the Lorenz number are also given.

| $T$ (K) | system | carrier concentration ($10^{12}$ cm$^{-2}$) | $S$ (μV/K) | $\sigma$ ($10^6$ S/m) | $S^2\sigma$ (W/mK$^2$) | $L$ ($10^{-8}$ WΩ/K$^2$) | $\kappa_e$ (W/mK) | $\kappa_{ph}$ (W/mK) | $ZT$ |
|---|---|---|---|---|---|---|---|---|---|
| 300 | x-p | 1.43 | 238 | 6.15 | 0.35 | 1.20 | 22.1 | 76.4 | 1.06 |
|  | x-n | 1.24 | −217 | 3.92 | 0.18 | 1.22 | 14.3 |  | 0.61 |
|  | y-p | 1.29 | 238 | 5.53 | 0.31 | 1.20 | 19.8 |  | 0.98 |
|  | y-n | 1.02 | −227 | 4.54 | 0.24 | 1.21 | 16.4 |  | 0.76 |
| 400 | x-p | 1.58 | 246 | 4.28 | 0.26 | 1.19 | 20.4 | 59.7 | 1.30 |
|  | x-n | 1.50 | −228 | 2.90 | 0.15 | 1.21 | 14.0 |  | 0.81 |
|  | y-p | 1.46 | 246 | 3.84 | 0.23 | 1.19 | 18.2 |  | 1.19 |
|  | y-n | 1.22 | −244 | 3.23 | 0.18 | 1.20 | 15.5 |  | 0.95 |
| 500 | x-p | 1.67 | 255 | 2.94 | 0.19 | 1.18 | 17.3 | 49.5 | 1.43 |
|  | x-n | 1.82 | −231 | 2.22 | 0.12 | 1.20 | 13.4 |  | 0.94 |
|  | y-p | 1.65 | 249 | 2.76 | 0.17 | 1.19 | 16.4 |  | 1.30 |
|  | y-n | 1.44 | −238 | 2.32 | 0.13 | 1.20 | 13.8 |  | 1.03 |
| 600 | x-p | 2.08 | 246 | 2.44 | 0.15 | 1.19 | 17.4 | 42.4 | 1.48 |
|  | x-n | 2.38 | −222 | 1.90 | 0.09 | 1.22 | 13.8 |  | 0.99 |
|  | y-p | 2.12 | 238 | 2.33 | 0.13 | 1.20 | 16.7 |  | 1.34 |
|  | y-n | 1.87 | −227 | 1.91 | 0.10 | 1.21 | 13.8 |  | 1.05 |
| 700 | x-p | 2.51 | 234 | 2.06 | 0.11 | 1.20 | 17.3 | 37.1 | 1.46 |
|  | x-n | 3.23 | −204 | 1.76 | 0.07 | 1.24 | 15.3 |  | 0.98 |
|  | y-p | 2.68 | 224 | 2.04 | 0.10 | 1.21 | 17.3 |  | 1.31 |
|  | y-n | 2.67 | −204 | 1.79 | 0.07 | 1.24 | 15.5 |  | 0.99 |
| 800 | x-p | 3.31 | 214 | 1.96 | 0.09 | 1.22 | 19.2 | 33.0 | 1.38 |
|  | x-n | 4.36 | −183 | 1.69 | 0.06 | 1.27 | 17.2 |  | 0.91 |
|  | y-p | 3.43 | 206 | 1.89 | 0.08 | 1.23 | 18.6 |  | 1.24 |
|  | y-n | 3.56 | −183 | 1.67 | 0.06 | 1.27 | 17.0 |  | 0.89 |
| 900 | x-p | 4.25 | 195 | 1.88 | 0.07 | 1.27 | 21.1 | 29.7 | 1.27 |
|  | x-n | 5.75 | −164 | 1.64 | 0.04 | 1.31 | 19.4 |  | 0.81 |
|  | y-p | 4.58 | 186 | 1.86 | 0.06 | 1.27 | 21.2 |  | 1.13 |
|  | y-n | 4.94 | −160 | 1.65 | 0.04 | 1.33 | 19.7 |  | 0.77 |
| 1000 | x-p | 5.14 | 179 | 1.77 | 0.06 | 1.28 | 22.6 | 27.0 | 1.14 |
|  | x-n | 7.69 | −145 | 1.65 | 0.04 | 1.37 | 22.5 |  | 0.70 |
|  | y-p | 5.87 | 169 | 1.82 | 0.05 | 1.31 | 23.7 |  | 1.02 |
|  | y-n | 6.76 | −141 | 1.65 | 0.03 | 1.37 | 22.7 |  | 0.66 |

## 4. Summary

In summary, we have studied the thermoelectric properties of $\gamma$-graphyne via first-principles calculations combined with the Boltzmann transport equations for both electron and phonon. As the generally used DP theory does not consider the contribution from the optical phonon scattering, the carrier relaxation time in the present work is carefully treated within the framework of complete electron-phonon coupling. It is thus anticipated that our calculated *ZT* values could give a better prediction of the thermoelectric performance of $\gamma$-graphyne. The maximum *ZT* value appears at 600 K, which is 1.5 for the *p*-type system along the *x*-direction and 1.0 for the *n*-type system along the *y*-direction. The significantly superior thermoelectric performance of $\gamma$-graphyne compared with graphene is originated from the existence of the carbon-carbon triple bonds and the opening of the band gap. Our theoretical work suggests that good thermoelectric performance can be also achieved in previously unexpected carbon materials, which has very promising prospect containing the earth-abundant and environment-friendly elements.

## Acknowledgments

We thank financial support from the National Natural Science Foundation (grant No. 11574236 and 51172167) and the "973 Program" of China (Grant No. 2013CB632502).